\documentclass{article}
\usepackage{epsfig}
\begin{document}
\title{\bf Stability and anyonic behavior of systems with $M$-statistics}
\author{Marcelo R. Ubriaco\thanks{Electronic address:ubriaco@ltp.uprrp.edu}}
\date{Laboratory of Theoretical Physics\\Department of Physics\\University of Puerto Rico\\R\'{\i}o Piedras Campus\\
San Juan\\PR 00931, USA}

\maketitle
\begin{abstract}
Starting with the partition function $Z$ for systems with $M$-statistics, as proposed in \cite{WND}, we calculate
from the metric $g_{\alpha\eta}=\frac{\partial \ln Z}{\partial\beta^{\alpha}\beta^{\eta}}$ the scalar curvature $R$
in two and three dimensions.  Our results exhibit the details of the anyonic behavior as a function of the fugacity $z$
and the identical particle maximum occupancy number $M$. We also compare the stability of systems for $M>1$
with the fermionic ($M=1$), and bosonic ($M\rightarrow \infty$), cases. 

\end{abstract}

\section{Introduction}
Several years ago a new fractional statistics was  proposed \cite{WND} in arbitrary dimensions 
which allows a finite number of multi-occupancy states of a single state.  Since that a single quantum state
can be occupied  by $n\leq M$ identical particles, this formalism opens the
way to a new thermodynamics based on an extension of the Pauli exclusion principle. From a calculation
of the second virial coefficient for low values of the fugacity $z$ it was found in \cite{WND} that an ideal gas becomes bosonic
for $M>1$.  Also, a calculation of the heat capacity $C_V$ for high $z$ values showed that the system remains fermionic
for $M>1$. Certainly, these approximate  calculations show that a thermodynamic system that follows $M$-statistics exhibit anyonic behavior for low $z$ values but miss the details that could tell us how this behavior depends on the fugacity $z$ and the maximum occupancy $M$. In addition, one important question that one may address is related to  the stability of systems based on $M$-statistics as compared with fermionic ($M=1$) and bosonic ($M\rightarrow \infty$) cases.
 The answer to these questions can be given by performing a calculation of the thermodynamic curvature $R$, which in our case, is nothing else than the scalar curvature in the two-dimensional space defined by the parameters $\beta$ and $\gamma=-\beta\mu$. The idea of using geometry to study some properties of thermodynamic systems \cite{Ti}-\cite{AN} opened the way to the basic formalism of
 defining a metric in a two dimensional parameter space and calculate the corresponding scalar curvature
as a measure of the correlations strength of the system \cite{IJKK}-\cite{R}, with applications to classical and quantum gases
\cite{R1}\cite{NS}\cite{JM1}\cite{BH}, magnetic systems \cite{JM2}-\cite{JJK}, non-extensive statistical mechanics \cite{T-B}-\cite{O},
anyon gas \cite{MH1}-\cite{MH2}, fractional statistics \cite{MH3}, deformed boson and fermion systems \cite{MH4}, systems
with fractal distribution functions \cite{MRU1} and quantum group invariance \cite{MRU2}. Some of the basic results of this formalism is the relationship between the departure of the scalar curvature $R$ from the zero value and the stability of the system, and the fact that $R$ vanishes for a classical gas, $R>0$ ($R<0$) for a boson (fermion) gas and becomes singular at a critical point.
Here, in order to study the stability and anyonic behavior of systems with M-statistics we calculate the dependence  of the scalar curvature on the fugacity $z$ and the maximum occupation  number $M$. In Section \ref {M}  we briefly describe the formalism of $M$-statistics. In Section \ref{SC} we calculate the scalar curvature for $M$-statistics in two and three dimensions. In Section \ref{C} we discuss and summarize our results.
\section{Quantum M-statistics}\label{M}
In this section we briefly introduce the formalism of $M$-statistics, the details and  the comparison of $M$-statistics with fractional exclusion statistics can
be found in Ref. \cite{WND}. For a single species of particles the set of orthonormal eigenstates of the number operator $\hat{N}$ are denoted by $|j>$ where
$j=0,1,2...,M$ and
\begin{equation}\hat{N}|j>=j|j>.
\end{equation}
The annihilation and creation operators $a$ and $a^{\dagger}$ satisfy a set of relations
\begin{equation}
(a a^{\dagger}\pm a^{\dagger} a)|j>= (|f_{j+1}|^2\pm |f_j|^2)|j>,
\end{equation}
where
\begin{equation}
a|j>=f_j|j-1>\;\;\; ,\;\;\; a^{\dagger}|j>=f^*_{j+1}|j+1>.
\end{equation}
In particular, $f_0=0$ and $f_1=e^{i\alpha}$. It is clear that $a^j=a^{\dagger j}=0$ for all values of $j$ such that $j>M$. For multi-species of
particles the states are simply written as $|i j k...>$ with the corresponding number operators satisfying commutation relations $[\hat{N}_{i},\hat{N}_{j}]=0$ and the $a$ and $a^{\dagger}$ operators satisfying two possible commutation rules
\begin{eqnarray}
a_ia_j&-&e^{\pm i \frac{2\pi}{M+1}}a_ja_i=0,\nonumber\\
a^{\dagger}_ia_j&-&e^{\mp i \frac{2\pi}{M+1}}a_ja^{\dagger}_i=0\;\;\;\;i<j.
\end{eqnarray}
It is simple to see that the states are fermionic for $M=1$, and become bosonic for $M\rightarrow \infty$.
The partition function takes the simple form
\begin{equation}
Z=\prod_p\sum_{j=0}^{M}z^je^{-j\beta\epsilon(p)},\label{Z}
\end{equation}
In the thermodynamic limit, we obtain in arbitrary dimensions $D$
\begin{equation}
\ln Z=VA(D)\int_0^{\infty}dp p^{D-1}\ln \frac{1-z^{M+1}e^{-(M+1)\beta\epsilon(p)}}{1-ze^{-\beta\epsilon(p)}},\label{lnZ}
\end{equation}
with $\epsilon(p)=p^2/2m$ and the factor $A(D)=\frac{1}{2^{D-1}\pi^{D/2}\Gamma(D/2)}$.
\section{Scalar Curvature} \label{SC}
The metric $g_{ij}$ for a thermodynamic system is defined \cite{IJKK} as the second order term after expanding the information distance
$I(\rho(\beta^{i}),\rho(\beta^{i}+d\beta^{i}))=Tr\rho(\ln\rho(\beta^i)-\ln\rho(\beta^{i}+d\beta^{i}))$ 
 between two statistical close states $\rho(\beta^{i})$ and $\rho(\beta^{i}+d\beta^i)$  leading to
\begin{equation}
g_{ij}=-\left<\frac{\partial^2\ln\rho}{\partial\beta^{i}\partial\beta^j}\right>,
\end{equation}
where in our case the thermodynamic coordinates are $\beta^1=\beta$ and $\beta^2=\gamma\equiv -\beta\mu$.
For exponential distributions, $\rho=\frac{exp(-\sum_i\beta^iF^i)}{Z}$, the metric simply reduces to
\begin{equation}
g_{ij}=\frac{\partial^2\ln Z}{\partial\beta^i\partial\beta^j}.\label{g}
\end{equation}
From Eqs. (\ref{Z}) and (\ref{g}) it is simple to obtain that the metric tensor components are written in terms of the second moments of the energy and particle number 
as follows
\begin{eqnarray}
g_{11}&=&<H^2>-<H>^2,\nonumber\\
g_{22}&=&<N^2>-<N>^2,\nonumber\\
g_{12}&=&<NH>-<N><H>.
\end{eqnarray}
From the metric we obtain the scalar curvature $R$  from the definition
\begin{equation}
R=\frac{2}{det g}R_{1212},
\end{equation}
where $det g=g_{11}g_{22}-g_{12}g_{12}$ and due to the obvious identities like $\frac{\partial g_{ij}}{\partial\beta^j}\equiv g_{ij,j}=g_{jj,i}$ the non-vanishing part of the curvature tensor $R_{ijkl}$ is given in terms of the
Christoffel symbols $\Gamma_{ijk}=\frac{1}{2}g_{ij,k}$ as follows
\begin{equation}
R_{ijkl}=g^{mn}\left(\Gamma_{mil}\Gamma_{njk}-\Gamma_{mik}\Gamma_{njl}\right).
\end{equation}
The calculation of $R$ simply reduces to solve
$$ R=\frac{1}{2 (detg)^2}\left|\begin{array}{ccc} g_{11} & g_{22} & g_{12}  \\ g_{11,1} & g_{22,1} & g_{21,1} \\
g_{11,2} & g_{22,2}& g_{21,2} \end{array}\right|.$$
where $g_{ij,k}=\partial g_{ij}/\partial \beta^k$.
From Eq. (\ref{lnZ}) we find that the metric components can be  written
\begin{eqnarray}
g_{11}&=&\frac{V}{\beta^2\lambda^D\Gamma(D/2)}H_{\frac{D}{2}+1}\nonumber\\
g_{12}&=&\frac{V}{\beta\lambda^D\Gamma(D/2)}H_{\frac{D}{2}}\nonumber\\
g_{22}&=&\frac{V}{\lambda^D\Gamma(D/2)}H_{\frac{D}{2}-1},
\end{eqnarray}
where the function $H_{\eta}$ is the integral

\begin{equation}
H_{\eta}=\int_0^{\infty}dx x^{\eta}\left(\frac{f'_0}{f_0^2}-\frac{(M+1)f'_M}{f_M^2}\right),
\end{equation}
where the function $f_L=1-z^{L+1}e^{-(L+1)x}\;(L=0,1,...,M)$ and $f'_L=\partial f_L/\partial \gamma$.
The function $H_{\eta}$ satisfies the simple relation
\begin{equation}
\frac{\partial H_{\eta}}{\partial\gamma}=-\eta H_{\eta-1}.
\end{equation}
We obtain for the scalar curvature $R$ for arbitrary $D$ 
\begin{equation}
R=\frac{\lambda^D\Gamma(\frac{D}{2})}{2V}\left[\frac{(1+\frac{D}{2})H^2_{\frac{D}{2}}H_{\frac{D}{2}-1}+(\frac{D}{2}-1)H_{\frac{D}{2}+1}
H_{\frac{D}{2}}H_{\frac{D}{2}-2}-
D H_{\frac{D}{2}+1}H^2_{\frac{D}{2}-1}}
{(H_{\frac{D}{2}+1}H_{\frac{D}{2}-1}-H^2_{\frac{D}{2}})^2}\right].\label{r}
\end{equation}
\\
Figures 1 and 2 show the results of a numerical calculation of Equation (\ref{r}) for the scalar curvature in $D=3$. Figure 1 shows the dependence of $R$ on the fugacity $z$ for the values $M=1,5,25$. For $M=5$ the system is bosonic 
for very low values of $z$ and becomes fermionic for $z>0.72$. The switch from bosonic to fermionic occurs at higher values of $z$ as $M$ increases. For large values of $z$ and  $M$,  these systems are more stable than the fermionic $M=1$ system.
There is a value of $z$ like $z\approx 0.7$ for $M=5$ and $z\approx 1$ for $M=25$ where the scalar curvature vanishes and thus $M$-systems mimic a classical gas behavior. Figure 2 shows the values of the scalar curvature $R$ as a function of 
$M\geq 1$ for the fugacity values $z=0.5,0.9,1.5,15$. For $z<1$, M-systems become bosonic as $M$ increases. For $z>1$ the behavior  is fermionic for all values of $M$. For any $z$ there is a value of $M>1$ after which the scalar curvature 
becomes constant. The instability increases at those values of $z\approx 1$ and larger values of $M$, as expected.
The behavior at $D=2$, shown in Figures 3 and 4,  is quite similar to the one  in three dimensions.  Figure 5 displays the scalar curvature for $0<z<1$ for the standard Bose-Einstein case $M\rightarrow\infty$, $M=25$ and $M=50$. The stability
is practically identical up to a value of $z$ where the Bose-Einstein system becomes more unstable approaching the onset of Bose-Einstein condensation and the $M$-systems are more stable and switching to  fermionic ($R<0$) at higher values of $z$.

\begin{figure}
\begin{center}
\epsfig{file= 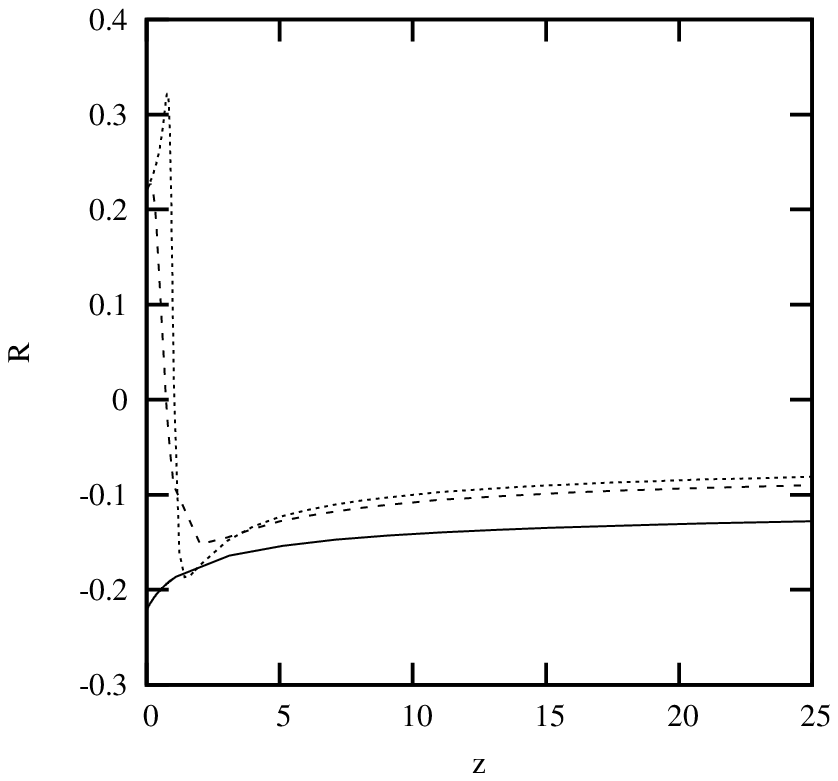,bbllx=50pt,bblly=120pt,bburx=430pt,bbury=250pt}
\end{center}
\caption[]{The scalar curvature $R$, in units of $\lambda^3V^{-1}$, as a function of the fugacity $z$ at $D=3$ and constant $\beta$ for the cases of $M=1$ (solid
line), $M=5$ (dashed line) and $M=25$ (dotted line).}
\end{figure}
\begin{figure}
\begin{center}
\epsfig{file= 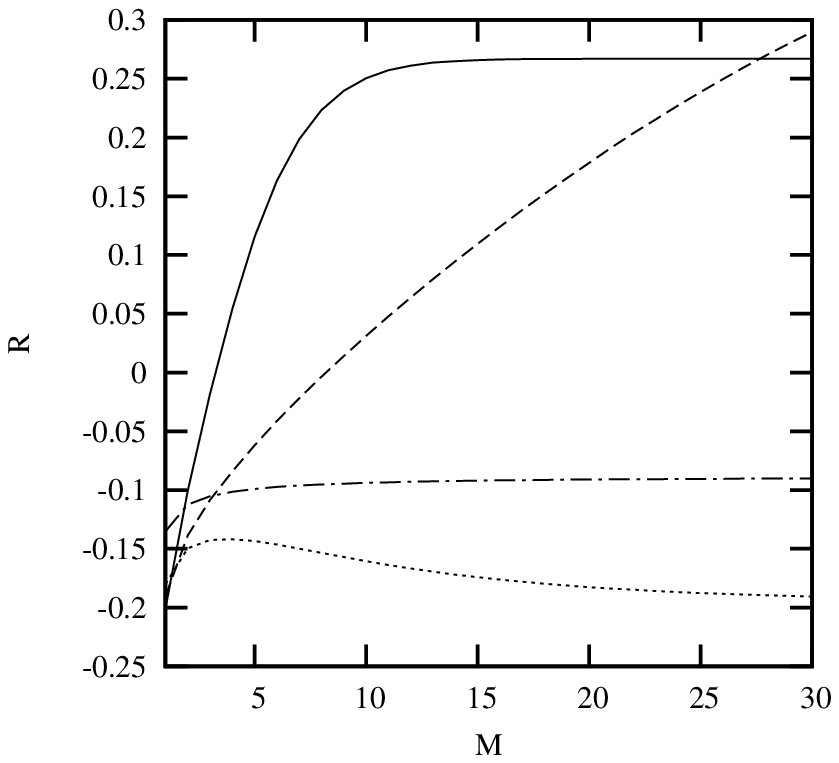,bbllx=50pt,bblly=120pt,bburx=430pt,bbury=250pt}
\end{center}
\caption[]{The scalar curvature $R$ at $D=3$, in units of $\lambda^3V^{-1}$, as a function of  $M$ and values for
the fugacity $z=0.5$ (solid line),   $z=0.9$ (dashed line), $z=1.5$ (dotted line) and $z=15$ (dashed-dotted line).}
\end{figure}
\begin{figure}
\begin{center}
\epsfig{file= 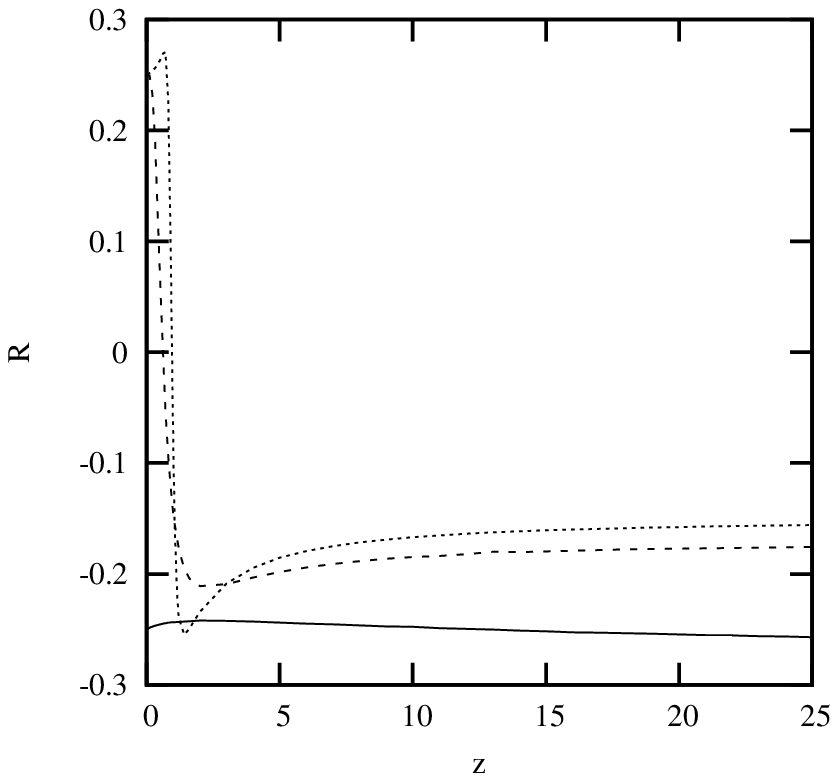,bbllx=50pt,bblly=120pt,bburx=430pt,bbury=250pt}
\end{center}
\caption[]{The scalar curvature $R$, in units of $\lambda^2A^{-1}$, as a function of the fugacity $z$  at $D=2$ and  constant $\beta$ for the cases of $M=1$ (solid
line), $M=5$ (dashed line) and $M=25$ (dotted line).}
\end{figure}
\begin{figure}
\begin{center}
\epsfig{file= 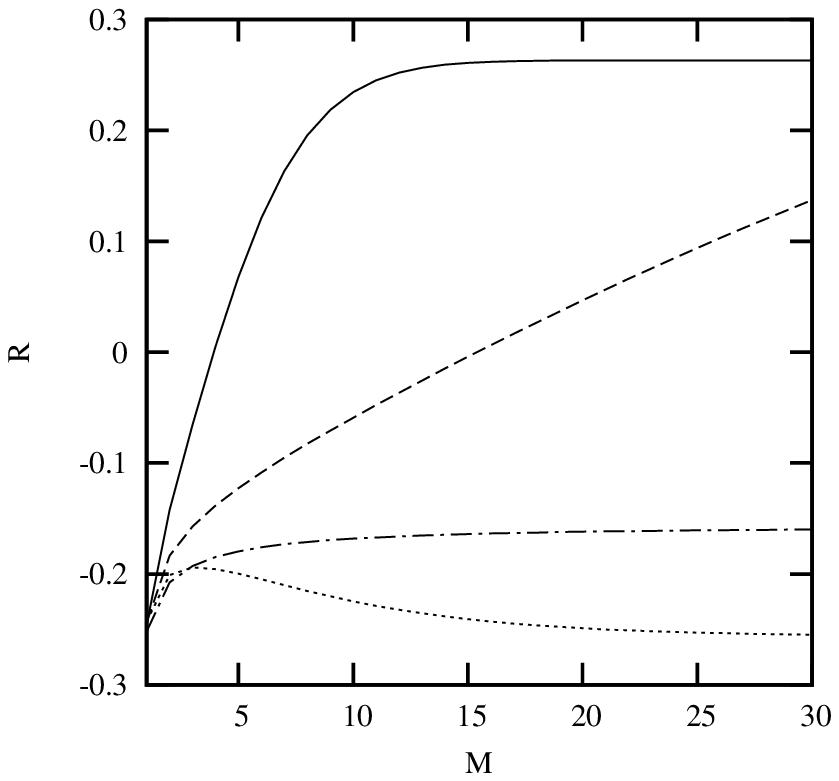,bbllx=50pt,bblly=120pt,bburx=430pt,bbury=250pt}
\end{center}
\caption[]{The scalar curvature $R$ at $D=2$, in units of $\lambda^2A^{-1}$, as a function of  $M$ and values for
the fugacity $z=0.5$ (solid line),   $z=0.9$ (dashed line), $z=1.5$ (dotted line)  and $z=15$ (dashed-dotted line).}
\end{figure}
\begin{figure}
\begin{center}
\epsfig{file= 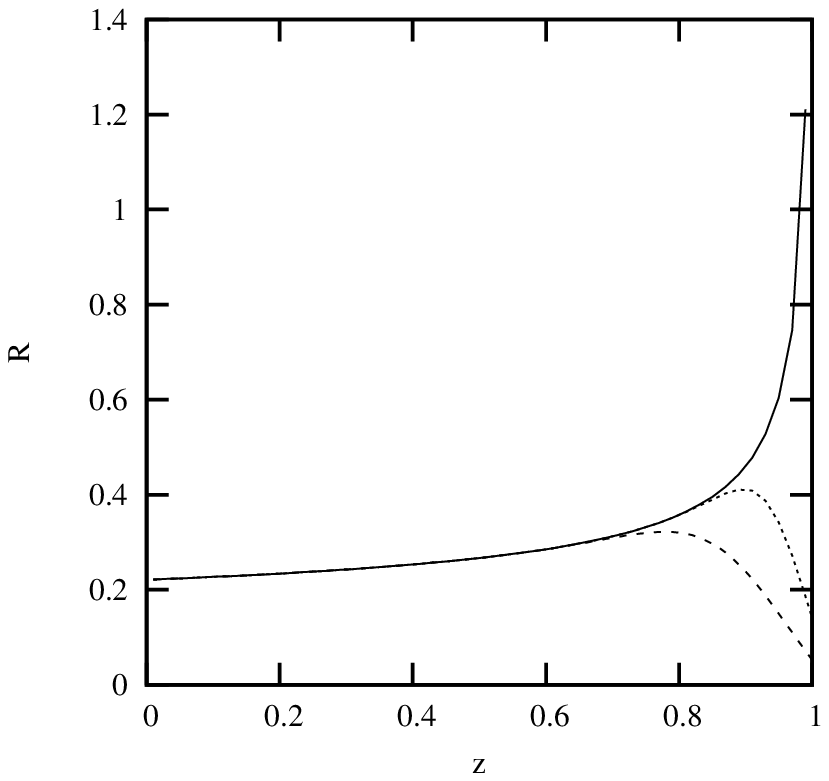,bbllx=50pt,bblly=120pt,bburx=430pt,bbury=250pt}
\end{center}
\caption[]{The scalar curvature $R$, in units of $\lambda^3V^{-1}$, as a function of the fugacity $z$, for $0<z<1$, at $D=3$ and constant $\beta$ for the cases a
Bose-Einstein gas $(M\rightarrow \infty)$ (solid
line), $M=25$ (dashed line) and $M=50$ (dotted line).}
\end{figure}
\section{Conclusions} \label{C}
In this manuscript we have calculated the scalar curvature for systems following $M$-statistics in two and three dimensions. Since the scalar curvature
$R$ is positive (negative) for bosons (fermions) our results give us a detailed picture about the anyonic behavior 
than a calculation of the virial coefficients could provide. In addition, the values of the scalar curvature tell us about the correlations and thus the stability of the system. A particular feature  of $M$-systems is that their behavior is very similar in two and three dimensions.  For those values $0<z<1$ these systems are bosonic and more stable than the 
standard Bose-Einstein case, and become fermionic at higher values of $z$. The change from bosonic to fermionic occurs at higher values of $z$ as the maximum occupation number $M$ increases. Therefore, $M$-statistics provides a different
framework, other than quantum group invariant systems \cite{MRU2}, to study anyonic behavior if ever observed in either two or three dimensions.
\section{Acknowledment}
I thank Y. J. Ng for reading the manuscript and useful comments.

\end{document}